\newcommand{\rem}[1]{}
\documentclass{amsart}
\usepackage{amsfonts,amssymb,amsmath,amsthm,mathrsfs}
\usepackage{url}
\usepackage{graphicx}
\newtheorem{thrm}{Theorem}[section]

\newtheorem{prop}[thrm]{Proposition}

\theoremstyle{definition}

\begin{document}
\author[C. A. Mantica and L. G. Molinari]{Carlo Alberto Mantica and Luca Guido Molinari}
\address{C.~A.~Mantica (ORCID: 0000-0001-5638-8655) and L.~G.~Molinari (corresponding author, ORCID:
0000-0002-5023-787X): 
Physics Department Aldo Pontremoli,
Universit\`a degli Studi di Milano and I.N.F.N. sezione di Milano,
Via Celoria 16, 20133 Milano, Italy.}
\email{carlo.mantica@mi.infn.it, luca.molinari@mi.infn.it}
\begin{abstract}
We introduce Sinyukov-like tensors, a special kind of conformal Killing tensors. In Robertson-Walker space-times they have the perfect-fluid form and only depend on two constants and the scale factor. They are the candidate for the dark term
of the newly proposed Conformal Killing Gravity, by Harada. In addition to ordinary matter, the Friedmann equations 
contain a dark term and a $\Lambda$ term that parametrize the Sinyukov-like tensor.
The expression of $H(z)$ is tested on cosmological data based on cosmic chronometers CC or including baryon acoustic oscillations BAO.  There is a large incertitude in $\Omega_\Lambda$ and  $\Omega_{dark}$ that may become negative, but their sum
is close to $\Omega_\Lambda$ of $\Lambda$CDM. In any case, there is a future singularity, that is a big rip for all $\Omega$s positive.
We solve the equation for the evolution, in linear approximation, of the density contrast in matter-dominated universe. The dark sector and the $\Lambda $ term give no significative deviation from $\Lambda$CDM and GR results. 
\end{abstract}
\title[]{Conformal Killing cosmology\\ 
Geometry, dark sector,\\
 growth of structures, and big rip}
\date{25 may 2024}
\maketitle

\section{Introduction}
Since the observation of supernovae SnIa by Riess and Perlmutter in 1998-99, dark energy stands as an unsolved theoretical problem. The reintroduction of the $\Lambda$ term in the Friedmann equations successfully accounted for a major accordance with observational cosmology, together with the cold dark matter (CDM) assumption.
Nevertheless the puzzle remains, with discrepancies between theory and data. Several modifications of the Standard Model 
have been proposed, in two mainstreams. The first one modifies the energy-momentum source with new fields
(see the review by Copeland, \cite{Copeland06}): phantom field \cite{Caldwell02}, quintessence \cite{Ratra88}, Chaplygin gas
\cite{Kamenshick01}, and others.
The second class modifies geometry (see \cite{Capozziello11,Saridakis21}) and contains popular models such as $f(R)$ \cite{Sotiriu10,Nojiri17}, $f(G)$ \cite{Nojiri05a}, $f(T)$ \cite{Cai16}, mimetic gravity \cite{Chamseddine13}, and others.
A third route is explored in \cite{Odintsov23}, with axion dark matter and dark energy resulting from modified gravity with
$f(R)=\exp(-\beta R)$.

In 2023 Junpei Harada \cite{Harada23a} introduced a modified
theory of gravity to explain the present accelerated phase
of the universe without the explicit introduction of dark energy.
It is based on the following field equations in $4$ dimensions:
\begin{align}
&\nabla_{j}R_{kl}+\nabla_{k}R_{jl}+\nabla_{l}R_{jk}-\tfrac{1}{3}(g_{kl}\nabla_{j}R+g_{jl}\nabla_{k}R+g_{jk}\nabla_{l}R)
\label{eq:Killing Harada-1}\\
&\quad=\nabla_{j}T_{kl}+\nabla_{k}T_{jl}+\nabla_{l}T_{jk}-\tfrac{1}{6}(g_{kl}\nabla_{j}T+g_{jl}\nabla_{k}T+g_{jk}\nabla_{l}T)
\nonumber
\end{align}
They are manifestly third order in the derivatives of the metric tensor and extend GR: a solution of the Einstein equations $R_{kl}-\frac{1}{2} Rg_{kl} =T_{kl}$ is a solutions of the new equations. 
In static spherical symmetry Harada obtained the Schwarzschild solution with new terms
that are significant at large distances, $-g_{00} = 1/g_{rr} =  1-\frac{2M}{r} -\frac{\Lambda}{3}r^2 - \frac{\lambda}{5}r^4 $. 
He then derived an equation for the scale factor in RW space-times. In a matter-dominated universe, the explicit solution shows a transition from decelerating to accelerating expansion without introducing dark sources in the equations; $\Lambda $
results as an integration constant.

Shortly after, we published a parametrization showing that Harada's equations
are equivalent to the Einstein equations modified by a supplemental conformal Killing tensor that is also 
divergence-free \cite{Mantica 23 a}:
\begin{align}
&R_{kl}-\tfrac{1}{2}Rg_{kl}=T_{kl}+K_{kl}\\
&\nabla_{j}K_{kl}+\nabla_{k}K_{jl}+\nabla_{l}K_{jk}=\tfrac{1}{6}(g_{kl}\nabla_{j}K+g_{jl}\nabla_{k}K+g_{jk}\nabla_{l}K)
\label{eq:Conformal Killing mantica-1}
\end{align}
For this reason the theory was named Conformal Killing Gravity (CKG).

The reformulation makes the extension of GR explicit through the Killing term, that satisfies $\nabla^k K_{kl}=0$ and poses as a natural candidate for the energy-momentum of the dark sector. In this direction we proved that a spacetime is generalized Robertson-Walker (GRW) if and only if it admits a conformal Killing tensor with the perfect-fluid form and with closed velocity vector. The condition of conformal flatness restricts the spacetime to RW. \\
We wrote the two Friedmann equations for the dark pressure and energy density, recovered Harada's solution for the
scale factor $a(t)$ in matter-dominated universe, and discussed a toy model. Both cases lead to the conclusion that the dark fluid determines the late time evolution of the scale parameter, with equation of state $p_D/\mu_D \to - \frac{5}{3}$ (phantom energy).

The cosmological scenarios of CKG were further explored by Harada in \cite{Harada23b} for various values of $\Omega_m$ in the evolution equation of the Hubble parameter 
\begin{align}
\frac{H(z)}{1+z} = H_0 \left [\Omega_m (1+z)+\Omega_r (1+z)^2 +\Omega_k +\frac{\Omega_\Lambda}{(1+z)^2} +\frac{\Omega_D}{ (1+z)^4} \right ]^{1/2} \label{HARADAH}
\end{align}
and compared with $\Lambda$CDM. If the effective dark energy $\Omega_D= 1-\Omega_m -\Omega_\Lambda $ (radiation and
curvature are neglected) is present in moderate amount, CKG holds the potential to resolve the Hubble tension. This 
remains true if $\Omega_D$ is dominant in the total energy budget with $\Lambda=0$. 

Solutions with $T_{kj}=0$ were obtained by Cl\'ement and Noucier \cite{Clement24} in a variety of cases
including singularity-free eternal cosmologies and universes evolving simmetrically in finite time from big-bang to big-crunch.  

The static spherical solution was retaken by Alan Barnes \cite{Barnes23a} to include the case $g_{rr}g_{00}\neq -1$, and raising the
interesting question of the validity of Birkhoff's theorem (spherical vacuum GR solutions are static).

Junior, Lobo and Rodrigues \cite{Junior24} studied black hole solutions in CKG coupled to nonlinear electrodynamics with
canonical or phantom-like scalar fields. 
They found generalizations of the Schwarzschild--Reissner-Nordstr\"om--AdS solutions that extend the class of known regular black hole solutions. In \cite{Junior24b} they explore black bounce solutions, that extend
Bardeen-type and Simpson-Visser geometries, in CKG coupled to NLED and scalar fields.\\
Solutions of CKG linearly coupled to Maxwell fields were obtained by Barnes \cite{Barnes23b} as powers series. The general closed-form solutions were then given by Cl\'ement and Nouicer \cite{Clement24}. 
Very recently Barnes obtained the most general pp-wave solutions in CKG, and their plane-wave specialisation
\cite{Barnes24} and claimed the unpublished result about conditions for the validity of Birkhoff's theorem in non-static
spherically symmetric vacuum metrics.

In this paper we investigate conformal Killing gravity in RW spacetime. 
To be specific, the conformal Killing tensor
\eqref{eq:Conformal Killing mantica-1} is perfect fluid and Sinyukov-like. In this case the geometry is wholly determined 
by two integration constants that parametrize the dark fluid. Their numerical values are fixed by the knowledge of 
$\Omega_\Lambda$ and $\Omega_D$ in the evolution \eqref{HARADAH} of $H(z)$, for which data
are still limited.

In section 2 we prove that a perfect fluid tensor $K_{ij}=Ag_{ij}+Bu_i u_j$ is a conformal Killing tensor 
if the velocity is shear-free. The fluid is conserved, $\nabla^i K_{ij}=0$, if the velocity is also vorticity-free. If the velocity is also acceleration-free the perfect fluid tensor is of special type: we name it Sinyukov-like. Conformal Killing gravity in
generalized Robertson-Walker space-times is thus provided by Sinyukov tensors, presented in section 3. In the RW setting the Ricci tensor is perfect fluid and the Sinyukov tensor is fixed by the scale parameter and two constants.\\
In section 4 we write the analogues of the Friedmann equations: the covariant approach is particularly effective for this. We give the formulae for $H(z)$, the deceleration $q(z)$ and the look-back time. 
In section 5, restricting to $\Omega_m+\Omega_\Lambda +\Omega_D=1$, a preliminary fit shows a qualitative difference if the dataset includes BAO or not. With exclusion the result is $\Omega_D>0$, allowing for an interpretation of dark energy density, and the occurrence of a ``big rip" future singularity. With inclusion it is $\Omega_D<0$: a different future singularity occurs, with 
$\dot a=0$, and the dark energy density remains positive for most of the universe lifetime.
In both cases $\Omega_m$ (matter) is stable and close to the $\Lambda$CDM value.\\
In section 6 we obtain the equation for the evolution of the density contrast in CKG (Sinyukov) cosmology
in matter dominance, with the $\Lambda $ term. It is exactly solved in the three relevant cases: \\
$\bullet$  $\Omega_m+\Omega_\Lambda =1$ ($\Lambda$CDM, $\Omega_D$ negligible). The evolution is given in terms  of a hypergeometric function and, for high $z$, it follows the GR solution $\Lambda =0$.\\
$\bullet$ $\Omega_m+\Omega_D=1$, $\Omega_\Lambda =0$: the growing mode is the same as in GR.\\
$\bullet$ $\Omega_m+\Omega_\Lambda+\Omega_D=1$: the solution, again, does not deviate 
from $\Lambda$CDM.

Notation: $i,j,k,\ldots =0,1,2,\ldots $; $\mu,\nu,\ldots=1,2,\ldots$ The dot derivative is $\dot{H}=u^{k}\nabla_{k}H$,
i.e. the time derivative in the comoving frame ($u^{0}=1$, $u^\mu=0$).

\section{Perfect fluid conformal Killing tensors}\label{SECTCKT}
In this section we study conformal Killing tensors that are perfect fluid tensors, and the restrictions they 
pose on the space-time.\\
In dimension $n$, a symmetric tensor $K_{ij}$ is a conformal Killing tensor (CKT) if  \cite{Rani 03} 
\begin{gather}
\nabla_i K_{jl}+\nabla_j K_{li}+\nabla_l K_{ij}=\eta_i g_{jl}+\eta_j g_{li}+\eta_l g_{ij} \label{CKT}\\
\eta_i=\frac{\nabla_i K + 2 \nabla_j K^j{}_i }{n+2} \label{ETAVECTOR}
\end{gather}
$\eta_{i}$ is the associated conformal vector, $K=g^{ij}K_{ij}$ is the trace, $\nabla_j K^j{}_i$ is the divergence
of the tensor. The CKT tensor is divergence-free if and only if: 
\begin{align}
\eta_i =\frac{\nabla_i K}{n+2}    \label{DIVFREE}
\end{align}
%
A perfect fluid tensor is characterized by a velocity vector field
$u_p u^p=-1$ and scalar fields $A$ and $B\neq 0$: 
\begin{equation}
K_{jk}=
Ag_{jk}+Bu_{j}u_{k}\label{eq:perfect fluid tensor}
\end{equation}
We recall that a velocity field has the canonical decomposition
\begin{equation}
\nabla_i u_j = H(g_{ij}+u_i u_j )-u_i \dot u_j+\omega_{ij} + \sigma_{ij} \label{CANONICAL}
\end{equation}
$H=\frac{1}{n-1}\nabla_ru^r $ is the expansion parameter, $\dot u^i =u^k\nabla_k u_i$ is the acceleration, $\omega_{ij}=-\omega_{ji}$ is the vorticity, $\sigma_{ij}=\sigma_{ji}$ is the shear, with $g^{ij}\sigma_{ij}=0$, $\omega_{ij}u^j =\sigma_{ij}u^j=0$.

The following theorem parallels statements in \cite{Sharma10} and \cite{Coll06}. A proof, simpler in our view and in tensor index notation, is given in Appendix 1:
\begin{thrm}\label{PFCKT}
The perfect fluid tensor (\ref{eq:perfect fluid tensor}) Coonformal Killing 
if and only if: \\
1) the velocity is shear-free,\\
2) the expansion parameter $H$ and the scalar $B$ satisfy
\begin{align}
 & H =\frac{\dot B}{2B} \label{expansion}\\
 & \nabla_j B=-\dot B u_j+2B\dot u_j.  \label{nablab}
\end{align}
Then the conformal vector is 
\begin{align}
\eta_j = \nabla_j A+\dot B u_j \label{CKVPF}
\end{align}
The scalar field $A$ is unconstrained.
\end{thrm}
Now we impose that the perfect fluid CKT is divergence-free.\\
The condition (\ref{DIVFREE})  
is $(n+2)\eta_j =\nabla_j K$ where $K=nA-B$. At the same time \eqref{CKVPF} holds. 
Then: 
\begin{align}
\dot B u_j= - \nabla_j \frac{2A+B}{n+2}  \label{dotBu}
\end{align}
The conformal vector becomes $\eta_j = - \frac{1}{2} (n-1)\dot B u_j - B\dot u_j$ and eq. \eqref{CKVPF} gives 
$$ \nabla_j A = - \frac{1}{2} (n+1)\dot B u_j - B\dot u_j $$
We distinguish two cases:\\
$\bullet $ $\dot B=0$: $\eta_j = \nabla_j A$ and $2A+B$ is a number. Since eq.\eqref{nablab} simplifies to 
$\nabla_j B = 2B \dot u_j$ the acceleration is closed. 
If also $\omega_{ij}=0$ then the spacetime is static: 
$$\nabla_i u_j =-u_i\dot u_j, \qquad \nabla_i\dot u_j = \nabla_j \dot u_i $$
$\bullet $ $\dot B\neq 0$: eq.\eqref{dotBu} shows that $u_i$ is hypersurface orthogonal. This implies that the vorticity is zero: $\omega_{ij}=0$. Then
$\nabla_i u_j = H(g_{ij}+u_i u_j) - u_i \dot u_j$.
\begin{prop}
If $\dot B\neq 0$ then either the acceleration $\dot u_j$ is not closed, or $\dot u_j=0$.
\begin{proof}
Suppose that the acceleration is closed: $ \nabla_i\dot u_j = \nabla_j \dot u_i$. 
Evaluate:
$0=\nabla_i\nabla_jB-\nabla_j\nabla_i B = -u_j\nabla_i\dot B +u_i\nabla_j \dot B -\dot B (u_i\dot u_j -\dot u_i u_j)$. This 
 implies $\nabla_j \dot B = -\ddot B u_j -\dot B \dot u_j$. Next evaluate: $0=\nabla_i\nabla_j A-\nabla_j\nabla_i A = (n+2)\dot B (u_i\dot u_j - u_j \dot u_i)$, which gives $\dot B \dot u_j =0$.
\end{proof}
\end{prop}

Hereafter we choose $\dot B\neq 0$ and  $\dot u_j=0$. This specializes the theory of conserved perfect-fluid CKT tensors to spacetimes of cosmology. \\
A Generalized Robertson Walker (GRW) spacetime has the warped metric
\begin{align}
ds^{2}=-dt^{2}+a(t)^{2}g_{\mu\nu}^\star ({\bf x})dx^{\mu}dx^{\nu},\label{eq:1}
\end{align}
where $g_{\mu\nu}^\star ({\bf x})$ is the Riemannian metric of a spacelike
hypersurface, and $a(t)$ is the scale factor. 
A covariant characterization is the existence of
a unit time-like vector field, $u_k u^k=-1$, such that (see \cite{Mantica 17,Capozziello 22}):
\begin{align}
&\nabla_j u_k =H (g_{jk}+u_j u_k),\label{eq:torse-forming}\\
& \nabla_j H = -\dot H u_j \label{dotH}
\end{align}
Condition \eqref{dotH} is equivalent to requiring that $u_j$ is an eigenvector of the Ricci tensor
$R_{ij} u^j =\xi u_i$. The eigenvalue is $\xi =(n-1)(H^2 + \dot H)$. \\
In cosmology $H$ is the Hubble parameter and
$\xi $ is related to the acceleration:
\begin{align}
H= \frac{\dot a}{a}, \qquad  \xi = (n-1)\frac{\ddot a}{a} 
\end{align}
The Ricci tensor and the scalar curvature in a GRW spacetime are
\begin{align}
&R_{kl}=\frac{R-n\xi}{n-1}u_{l}u_{k}+\frac{R-\xi}{n-1}g_{kl}-(n-2)E_{kl}.\label{eq:2.3 Ricci GRW}\\
&R=\frac{R^\star}{a^2} + (n-1)(n-2) H^2 +2 \xi     \label{eq:2.9 scalar R}
\end{align}
where $E_{kl}=u^{j}u^{m}C_{jklm}$ is the electric part of the Weyl curvature tensor $C_{jklm}$ and
$R^\star$ is the curvature of the spacelike hypersurface.

\begin{prop}
If $K_{ij}=Ag_{ij}+ Bu_i u_j$ is a divergence-free CKT with $\dot u_j=0$, then the spacetime is GRW.
\begin{proof}
By hypothesis  \eqref{eq:torse-forming} holds, we only need showing eq.\eqref{dotH}. With
$H=\dot B/(2B)$ and $\nabla_j B= -\dot B u_j$ let's evaluate:
$$\nabla_j \dot B = \nabla_j (u^k\nabla_k B) = (\nabla_j u^k)(-\dot B u_k) + u^k \nabla_j \nabla_k B = u^k \nabla_k
 \nabla_j B = - u_j \ddot B$$
Therefore $\nabla_j H = -  [\ddot B/(2B) + (\dot B)^2/(2B^2)]u_j =-\dot H u_j$.
\end{proof}
\end{prop}

If $\dot u_j=0$, the perfect fluid divergence-free CKT tensor lives in a GRW spacetime and has a special property:
\begin{align}
\nabla_i K_{jl} &= g_{jl}\nabla_i A + \nabla_i (B u_j u_l) \nonumber \\
&= -\frac{1}{2}(n+1)\dot B u_i g_{jl} -\dot B u_i u_j u_l + B H(2 u_i u_j u_l + g_{ij} u_l + g_{il} u_j) \nonumber \\
&= [-(n+1) BH u_i] g_{jl}  + [BHu_j] g_{li}  + [BH u_l] g_{ij} \label{SINY}
\end{align}
This motivates the definition in the next section.

\section{Sinyukov-like tensors and GRW spacetimes}
We consider symmetric tensors satisfying the relation 
\begin{align}
\nabla_j K_{kl} = a_j g_{kl}+b_k g_{jl}+b_l g_{jk}        \label{eq:Sinyukov}
\end{align}
and call them ``Sinyukov-like'' tensors \cite{Mantica12,Step12}. When $K_{kl}$ is the Ricci tensor
we recover the characterization of Sinyukov manifolds, investigated for example in \cite{Formella 95}.

Contractions with $g^{jk}$ and $g^{kl}$ give  $a_j$ and $b_j$
in terms of $K=g^{ij}K_{ij}$ and  the divergence of the tensor. Thus, explicitly: 
\begin{align}
\nabla_{j}K_{kl}=&\frac{(n+1)\nabla_{j}K-2\nabla^{p}K_{pj}}{(n+2)(n-1)}g_{kl} \label{eq:eq5}\\
&+\frac{n\nabla^{p}K_{pk}-\nabla_{k}K}{(n+2)(n-1)}g_{jl}+\frac{n\nabla^{p}K_{pl}-\nabla_{l}K}{(n+2)(n-1)}g_{jk}\nonumber
\end{align}
The cyclic sum of \eqref{eq:eq5} shows that a Sinyukov-like tensor is a CKT. 

In the end of Section \ref{SECTCKT} we showed that a perfect fluid conformal Killing tensor $K_{ij}=A g_{ij}+ B u_i u_j$ that is divergence-free and acceleration-free i.e. 
 \begin{align}
& \nabla_i u_j = H (g_{ij}+u_iu_j)\\
&H=\frac{\dot B}{2B}, \quad \nabla_j B = -\dot B u_j , \quad \nabla_j A= -\frac{n+1}{2}\dot B u_j \label{EQS123}
\end{align}
 is a Sinyukov-like tensor, and the spacetime is a GRW.\\
Now we easily prove the opposite: in a GRW spacetime characterized by the velocity field $u_j$, a perfect fluid tensor 
$Ag_{ij} + Bu_i u_j$  with $A$ and $B$ satisfying \eqref{EQS123} is Sinyukov-like and divergence-free.
\begin{proof}
The evaluation \eqref{SINY} proves that $K_{ij}$ is Sinyukov like. The contraction with $g^{ij}$ shows that it is also
divergence-free.
\end{proof}
We then conclude:
\begin{thrm} A spacetime is GRW  if and only if it admits a divergence-free, acceleration free, perfect fluid Sinyukov-like tensor. 
\end{thrm}

Equations \eqref{EQS123} can be solved in terms of the scale function $a(t)$.\\ 
The first one, $H=\dot a/a = \dot B/(2B)$, has solution $B(t) = \frac{1}{n-1}C a^2(t)$ where $C$ is constant. The second and third equations give 
$A(t) =\frac{1}{2}(n+1) B(t) - \Lambda $ where $\Lambda $ is another constant. Therefore, the final form of the Sinyukov-like divergence-free tensor is only parametrized by the scale function $a(t)$ and the constants $C$ and $\Lambda$:
\begin{align}
K_{kl}= g_{kl} \left[ \frac{1}{2}\frac{n+1}{n-1} Ca^2 - \Lambda \right ] + u_k u_l \frac{Ca^2}{n-1} \label{eq:Conformal Killing GRW}
\end{align}
Hereafter we set the spacetime dimension $n=4$ in all the equations.

\section{Conformal Killing gravity in RW spacetimes}
A GRW spacetime is RW whenever the Weyl tensor is zero. With $E_{kl}=0$, the Ricci tensor \eqref{eq:2.3 Ricci GRW} has the perfect-fluid form and the Riemann tensor is \cite{Mantica16}
\begin{align}
R_{jklm}=&\frac{1}{6}(R-2\xi) (g_{km}g_{jl}-g_{kl}g_{jm}) \label{eq:Riemann FRW}\\
&+\frac{1}{6}(R-4\xi)(g_{km}u_{j}u_{l}-g_{kl}u_{j}u_{m}-g_{jm}u_{k}u_{l}+g_{jl}u_{k}u_{m})  \nonumber
\end{align}

Consider the equation \eqref{eq:Conformal Killing mantica-1} of Conformal Killing Gravity in a RW spacetime with the Sinyukov-like perfect fluid term $K_{kl}$ describing the imposing {\em Dark sector} of gravity.\\
Eqs. \eqref{eq:2.3 Ricci GRW} and \eqref{eq:Conformal Killing GRW} give the perfect fluid 
energy-momentum tensor of ordinary matter with energy density $\mu $ and pressure $p$:
\begin{align}
T_{kl}&=(p+\mu)u_k u_l + p g_{kl} \nonumber\\
&=R_{kl}-\tfrac{1}{2}Rg_{kl}-K_{kl} \nonumber\\
&=-\tfrac{1}{6} (R+2\xi+5Ca^2-6\Lambda)g_{kl}+\tfrac{1}{3}\left (R-4\xi-Ca^2 \right )u_k u_l \label{eq:stress energy conformal killing}
\end{align}
After specifying $R$ with \eqref{eq:2.9 scalar R} and $\xi = 3(H^2+\dot H)$, the Friedmann equations for CKG with ordinary matter are:
\begin{align}
&\mu =\frac{R^{\star}}{2a^{2}}+3H^{2}+\frac{1}{2}Ca^{2}- \Lambda \label{KAPPAMU}\\
& p =-\frac{R^{\star}}{6a^{2}}-3H^{2}-2\dot{H}-\frac{5}{6}Ca^{2}+\Lambda
\label{eq:Friedmann conformal killing}
\end{align}
They reduce to the GR equations when $C=0$ and $\Lambda =0$.\\
We obtain 
\begin{align}
\mu+3p =-6\frac{\ddot{a}}{a}-2Ca^{2}+2\Lambda\label{eq:third friedmann}
\end{align}
In \cite{Mantica 23 a} we showed that \eqref{KAPPAMU}, \eqref{eq:Friedmann conformal killing} imply eq.(32) by Harada \cite{Harada23a}. The converse was shown by Clem\'ent and Noucier \cite{Clement24}.

Let ordinary matter be composed of dust with energy density $\mu_m$ and zero pressure, and radiation with EoS $p_r=\frac{1}{3}\mu_r$. Then $\mu=\mu_r+\mu_m$ and $p=p_r$. They are conserved independently. Dust: $0=\nabla_k(\mu_m u^k u_l) = (\dot \mu_m  + 3H\mu_m)u_l $. Radiation: $0=\nabla_k (\mu_r+p_r)u^ku_l +\nabla_l p_r=\frac{4}{3}(\dot \mu_r +3H\mu_r) u_l + \frac{1}{3}\nabla_l \mu_r$. The two equations give: $\mu_m =\mu_{m0}(\frac{a}{a_0})^{-3}$ and $\mu_r = \mu_{r0}(\frac{a}{a_0})^{-4}$ where $a_0=a(t_0)$ is the present-time scale.

The Friedmann equation \eqref{KAPPAMU} of CKG with radiation, matter, dark energy, $\Lambda$ term, 
and curvature term is
$$ \mu_{m0}\left (\frac{a}{a_0}\right )^{-3}+\mu_{r0}\left (\frac{a}{a_0}\right )^{-4} = \frac{R^{\star}}{2a^{2}}+3H^{2}+\frac{1}{2}Ca^{2}-\Lambda $$
Divide by $3H_0^2$ (the value at scale $a_0$) and obtain: 
\begin{align}
\left(\frac{H}{H_{0}}\right)^{2}=\Omega_{r}\left(\frac{a}{a_{0}}\right)^{-4} + \Omega_{m}\left(\frac{a}{a_{0}}\right)^{-3}+  
\Omega_{k}\left(\frac{a}{a_{0}}\right)^{-2} + \Omega_{\Lambda}+
\Omega_{D}\left(\frac{a}{a_{0}}\right)^2   \label{eq:ALL H}
\end{align}
$$\Omega_{m} = \frac{\mu_{m0}}{3H_0^2}, \; \Omega_{r} = \frac{\mu_{r0}}{3H_0^2},\;
\Omega_{k}=-\frac{R^{\star}}{6H_{0}^{2}a_{0}^{2}}, \; \Omega_{\Lambda}=\frac{\Lambda}{3H_{0}^{2}}, \; 
\Omega_{D}=-\frac{Ca_{0}^{2}}{6H_{0}^{2}} $$
Evidently, the cosmological parameters satisfy $\Omega_{m}+\Omega_{r}+\Omega_{k}+\Omega_{\Lambda}+\Omega_{D}=1$.\\
Since $H=\dot a/a$, equation \eqref{eq:ALL H} gives the time evolution of the scale function. 
In terms of the redshift $1+z= a_0/a $, it becomes eq.\eqref{HARADAH} by Harada:
\begin{align}
\left(\frac{H(z)}{H_0}\right)^2 = \Omega_r (1+z)^4 + \Omega_m (1+z)^3+  
\Omega_k(1+z)^2 + \Omega_\Lambda + \frac{\Omega_D}{(1+z)^2} \label{HSQ}
\end{align}
The ``lookback time'' of a photon emitted at time $t_e$ at scale $a_e = a(t_e)$, and received at 
present time $t_0$ at scale factor $a_0$, is:
\begin{align}
t_L = \int_{t_e}^{t_0} dt=\int_{a_e}^{a_0} \frac{da}{\dot a} 
=  \int_0^{z_e} \frac{dz}{(1+z)  H(z)}
\end{align}
The age of the Universe, if finite, is the limit $z\to \infty$ of $t_L$.\\
The future time-span of the Universe from now ($z=0$), is the integral
\begin{align}
\tau = \int_{z_f}^0 \frac{dz}{(1+z) H(z)} 
\end{align}
We obtain a bound for the future time-span, valid when  all $\Omega_j>0$, $z_f=-1$: 
\begin{align}
\tau H_0  \le \int_{-1}^0 \frac{dz}{\sqrt {\Omega_\Lambda (1+z)^2 + \Omega_D}} 
= \frac{1}{2\sqrt{\Omega_\Lambda}}
\log \frac{ \sqrt{\Omega_D+\Omega_\Lambda}+\sqrt{\Omega_\Lambda}}{\sqrt{\Omega_D+\Omega_\Lambda}-\sqrt{\Omega_\Lambda}} \label{ESTIMATE}
\end{align}

%

\section{CKG parameters from data fits for $H(z)$}
In the following we consider a flat space submanifold $\Omega_k=0$, and also $\Omega_r=0$. 
The smallness of $\Omega_r\approx 9.16\times 10^{-5}$ (neutrino corrected) compared to  $\Omega_{m}=0.31$ (Planck data)
determines a large value $1+z_m =\Omega_{m}/\Omega_{r}$ where all but matter contributions
are negligible. At decreasing redshifts the universe is in the matter-dominated era until $\Lambda $ and $D$ take over.

With three relevant terms it is:
\begin{align}
\left(\frac{H(z)}{H_0}\right)^2 = \Omega_{m}(1+z)^3+  \Omega_{\Lambda}+ \frac{\Omega_{D}}{(1+z)^2} \label{H2MLD}
\end{align}
If $\Omega_D>0$ the squared Hubble parameter diverges to $+\infty $ for  $z\to -1^-$ (future singularity).
If  $\Omega_D<0$ the requirement $H^2(z)\ge 0$ determines a critical value $H^2(z_c)=0$ where $\dot a=0$ at finite   
scale $a=a_0/(1+z_c)$.

\noindent
The deceleration parameter is $q=-\dfrac{a\ddot a}{\dot a^2} = - 1- \frac{\dot H}{H^2}$. With
$\dot{H}= - \frac{1}{2} \frac{dH^2}{dz}(z+1)$, it is 
\begin{align}
q(z) = \frac{\Omega_m (1+z)^5 -2\Omega_\Lambda (1+z)^2 -4\Omega_D}{2\Omega_m (1+z)^5 +2\Omega_\Lambda (1+z)^2 +2\Omega_D} \label{qz}
\end{align}

As a preliminary test we make a best fit for the parameters $\alpha_k$ in 
$$H(z) = \sqrt{\alpha_m (1+z)^3 + \alpha_\Lambda + \alpha_D (1+z)^{-2}}$$
with the available dataset  $(z_j, H_j, \sigma_j)$ for cosmic chronometers (CC, 32 points) and 
CC+BAO (58 points) taken from \cite{Moresco23,Tomasetti23,Moresco12,Sharov18,Qi23}. 
The fit is made with Mathematica, {\sf NonLinearFit} with weights.
\begin{center}
\begin{tabular}{|c || c | c || c | c || }
\hline
 {} & CC                     & CC    &  CC+BAO           &CC+BAO\\
{} & $\Lambda$CDM & CKG & $\Lambda$CDM & CKG  \\
\hline
$\alpha_m$            &1483$_{\pm 173}$ & 1586$_{\pm 303}$ & 1309$_{\pm 51}$ & 1262$_{\pm 63}$ \\
$\alpha_\Lambda$ &3163$_{\pm 540}$ &1877$_{\pm 3119}$ & 3567$_{\pm 193}$ & 4544$_{\pm 788}$   \\
$\alpha_D$            & --                            &1638$_{\pm 3919} $        & --  & -1688$_{\pm 1320}$                 \\
\hline
\end{tabular}
\end{center}
Next we evaluate $H_0 = H(0)$, $\Omega_k=\alpha_k/H_0^2$, and related values in the table:
\begin{center}
\begin{tabular}{| c || c |c || c | c || c | c |}
\hline
& CC & CC &  CC+BAO &CC+BAO\\
 & $\Lambda$CDM & CKG &  $\Lambda$CDM & CKG\\
\hline
$H_0$                       & 68.16 & 71.42     & 69.83 &64.17\\
$q_0$ & -0.516 &-0.854 &-0.549 & -0.130\\ 
$\Omega_m$             & 0.323 & 0.311  & 0.301 & \,0.306 \\
$\Omega_\Lambda$  & 0.677 & 0.368   &0.699 & \,1.103\\
$\Omega_D$             & --        & +0.321  & --         & -0.410\\
Age                            &13.59  & 13.47   & 13.92 &\,14.10 \\
\hline
\end{tabular}
\end{center}
$\bullet $ With CC data one notes that $\Omega_m$ is little changed by the presence of the Sinyukov terms.
The $\Lambda$CDM value of $\Omega_\Lambda$ splits into the sum of the two dark terms. \\
The $\Lambda$CDM and CKG fitting lines almost overlap in the past, and deviate for $z$ around zero.
The deviation is more marked in $q(z)$ towards the future: driven by the $D$ term the CKG universe accelerates more than the $\Lambda$CDM solution, driven by the $\Lambda$ term.\\
$\bullet$ With CC+BAO data, the significant feature is $\Omega_D<0$. This makes $H^2(z)$ vanish at $z_c=-0.407$ which is a limit value ($H^2<0$ in the interval $(-1,z_c)$).

Given the great uncertainty of the values $\alpha_\Lambda$ and $\alpha_D$, a fit with SnIa and 
new data with small $z$ are highly desirable.

\begin{figure}[h]
\begin{center}
\includegraphics*[width=12.5cm,clip=]{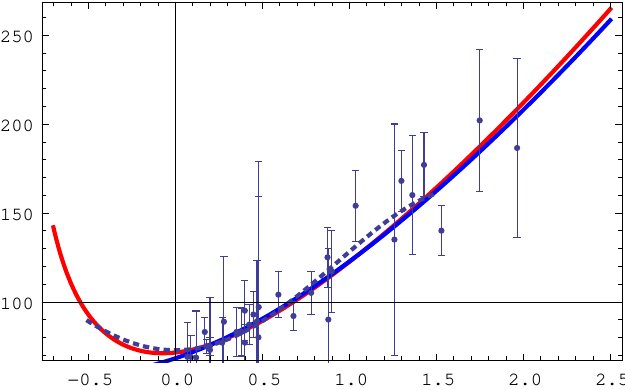}
\caption{Best fit for $H(z)$ for $\Lambda$CDM (blue) and CKT (red), with 32 experimental data with cosmic chronometers, including the recent point $z=1.26$ \cite{Tomasetti23}. The dotted line is the output of the genetic algorithm with CC data (31 points), eq.7 in \cite{Gangopadhyay23}, extrapolated out of $0<z<1.4$.}
\end{center}
\end{figure}

\begin{figure}[t]
\begin{center}
\includegraphics*[width=12.5cm,clip=]{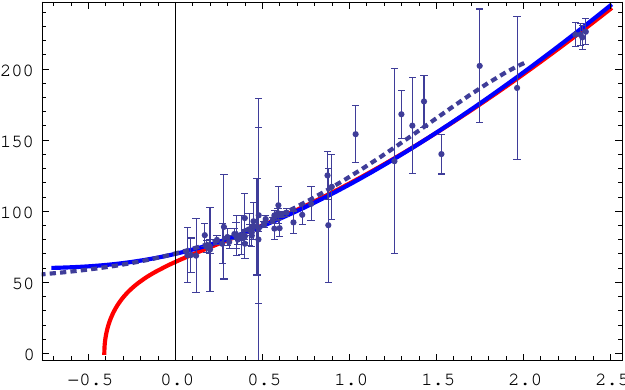}
\caption{Best fit for $H(z)$ for $\Lambda$CDM (blue) and CKT (red) and 58 CC+BAO data. The dotted line is the output of the genetic algorithm, eq.9 in \cite{Gangopadhyay23}  extrapolated out of $0<z<1.4$.}
\end{center}
\end{figure}

\begin{figure}[h]
\begin{center}
\includegraphics*[width=10cm,clip=]{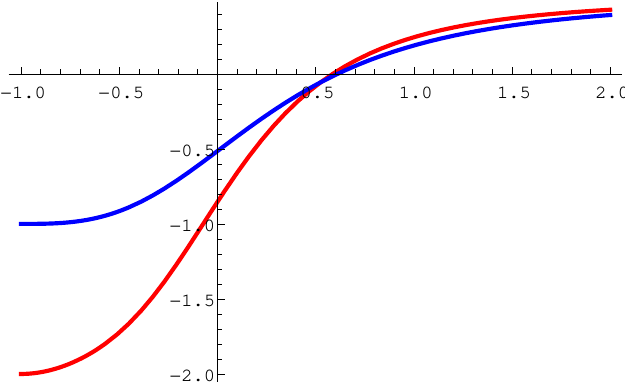}
\caption{The deceleration parameter $q(z)$, eq.\eqref{qz}, for $\Lambda$CDM (blue) and CKT (red) (data from CC only)}
\end{center}
\end{figure}

\noindent
The future time-span of the Universe from now ($z=0$), is the integral
\begin{align}
\tau = \int_{z_f}^0 \frac{dz}{(1+z) H(z)} = 
\begin{cases} 
21.0\,{\rm Gy} & \Omega_D>0, \quad z_f=-1\\ 
15.1 \,{\rm Gy} & \Omega_D<0,\quad z_f=-0.407 \end{cases}
\end{align}
The value $\tau$ for $\Omega_D>0$ is only slightly smaller than the estimate $\tau< 21Gy$  with eq.\eqref{ESTIMATE}: this signals the ongoing marginality of the matter term.\\
Differently from $\Lambda$CDM, $\tau $ is finite because 
of the occurrence of future singularities. Future singularities are discussed in \cite{Nojiri05,deHaro23}. \\
$\bullet$ $\Omega_D>0$ both $H(z)$ and $a(t)$ diverge at $z=-1$. The parameters of the Sinyukov tensor \eqref{eq:Conformal Killing GRW} may be interpreted as energy density and pressure of the dark perfect fluid:
\begin{align}
&\mu_D = -\frac{1}{2}Ca^2(t) + \Lambda = 3H_0^2 \left[ \frac{\Omega_D}{(1+z)^2} + \Omega_\Lambda \right ] \\
& p_D = \frac{5}{6}Ca^2(t)-\Lambda = -3H_0^2 \left[ \frac{5}{3} \frac{\Omega_D}{(1+z)^2}  +\Omega_\Lambda \right ]
\end{align}
$\mu_D>0$ and $p_D<0$.  As both diverge, the future singularity is a ``big rip" \cite{Caldwell03}. \\
$\bullet $ $\Omega_D<0$: $H(z)$ vanishes as a square root at $z(\tau)$ and $a(\tau)$ is finite. Given the values of the CC+BAO fit, for $z>-0.39$ the energy density is positive; for $z> - 0.21$ the pressure is negative. At $z(\tau)$ the pressure is finite and positive, while $\mu_D<0$. This future singularity is not in the scheme in \cite{Nojiri05}.
\begin{figure}[h]
\begin{center}
\includegraphics*[width=6cm,clip=]{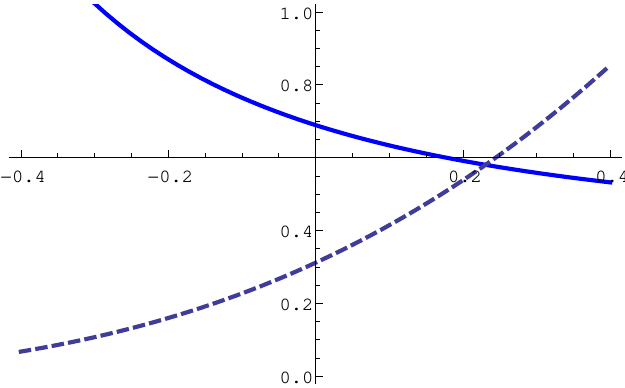}
\includegraphics*[width=6cm,clip=]{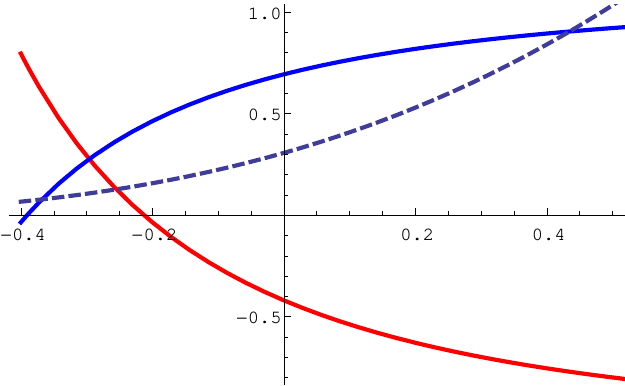}
\caption{The dark energy density $\mu_D(z)$ (blue), dark pressure $p_D(z)$ (red) and the matter energy density 
$\mu_m(z)$ (dashed) in units $3H_0^2$, with CC (left) and CC+BAO (right). For CC the pressure is negative and not shown. Note that the dark energy density $\mu_D$ is positive throughout almost the whole range. It turns negative close to
the future singularity with CC+BAO data.}
\end{center}
\end{figure}

\section{Growth of perturbations in CKG cosmology}

In this section we investigate the equations governing
the growth of perturbations. 
We follow the framework of spherical collapse illustrated by Abramo et al. \cite{Abramo07}. The procedure has been used in  several theories of extended gravity, such as mimetic gravity \cite{Farsi22}, energy momentum-squared gravity \cite{Farsi23}, time-dependent $\Lambda $ \cite{Silveira94}, modified torsion cosmology \cite{Usman23}, generalized Rastall gravity \cite{Ziaie20}.

We set the ordinary matter of the spatially flat FRW Universe ($R^\star =0$) to be a pressure-less dust, 
in presence of the $\Lambda $ and the dark terms.
The equations \eqref{KAPPAMU} and \eqref{eq:Friedmann conformal killing} give:
\begin{align}
 \mu_m = 3H^2 +\frac{C}{2}a^2 - \Lambda,  \label{PERT1}\\
 \frac{\ddot a}{a} = -\frac{\mu_m}{6} -\frac{C}{3}a^2 +\frac{\Lambda}{3}  \label{PERT2}
 \end{align}
Consider a local density perturbation that changes the background value $\mu_m$ to $\mu_m^c=\mu_m (1+\delta_m)$. 
In spherical symmetry, the mass content $\mu_m a^3$ in a spherical volume with radius $a^3$, fills a spherical volume with radius $ a_p^3$ with mass density $\mu_m^c$. From $\mu_m^c a_p^3=\mu_m a^3$ we infer 
$$a_p = a (1+\delta_m)^{-1/3} \approx a(1-\tfrac{1}{3}\delta_m) $$
The parameter $\delta_m =(\mu_m^c/\mu_m)-1$ is the ``density contrast'' \cite{Peebles}. 

Let the conservation rule for the perturbed sphere be
$\dot{\mu}_{m}^{c}+3h\mu_{m}^{c}=0$, where $h= \dot a_p/a_p$ is the ``local'' Hubble parameter. \\
While the background Universe evolves as \eqref{PERT2},
%
an analogous evolution for the perturbed cluster is assumed: 
\begin{equation}
\frac{\ddot{a}_p}{a_p}=-\frac{\mu_m^c}{6}-\frac{C}{3}a_p^2  +\frac{\Lambda}{3}.  \label{evolut cluster}
\end{equation}
In the linear approximation it is $\dot a_p= \dot a(1-\frac{1}{3}\delta_m)-\frac{1}{3}a\dot\delta_m $. Another dot-derivative gives
$\ddot a_p = \ddot a  (1-\frac{1}{3}\delta_m) -\frac{2}{3}\dot a\dot\delta_m -\frac{1}{3}a\ddot\delta_m$. Divide by $a_p$ and use the
evolution equations:
$$-\frac{\mu_m^c}{6}-\frac{C}{3}a_p^2 = -\frac{\mu_m}{6}-\frac{C}{3} a^2 -\frac{2}{3}H\dot\delta_m -\frac{1}{3}\ddot\delta_m $$
Simplify inserting $\mu_m^c=\mu_m (1+\delta_m)$ and $a_p^2=a^2(1-\frac{2}{3}\delta_m)$, use \eqref{PERT2} to specify $\mu_m$.
Then:
\begin{align}
\ddot\delta_m +  2H\dot\delta_m - \left[\frac{3}{2}H^2 - \frac{5C}{12}a^2 -\frac{\Lambda}{2}\right ] \delta_m =0  \label{eq:lineariz conts evolut}
\end{align}
Trade the dot derivative with the derivative with respect to the scale factor:
\begin{align*}
&\dot\delta_m = \frac{d\delta_m}{da} \dot a =  \frac{d\delta_m}{da} H a\\ 
&\ddot \delta_m = \frac{d^2\delta_m}{da^2}\dot a^2 + \frac{d\delta_m}{da} \ddot a =  \frac{d^2\delta_m}{da^2}H^2a^2 - 
\frac{d\delta_m}{da} a \left[\frac{H^2}{2}+\frac{5}{12}C a^2 -\frac{\Lambda}{2}\right]
\end{align*}
and obtain:
\begin{align}
\frac{d^2\delta_m}{da^2} a^2+ 
a\frac{d\delta_m}{da} \left[\frac{3}{2}- \frac{5C}{12H^2}a^2 +\frac{\Lambda}{2H^2}\right] 
 - \left[\frac{3}{2} - \frac{5C}{12 H^2} a^2 -\frac{\Lambda}{2H^2} \right ] \delta_m =0  
\end{align}
The equation may be written in red-shift
space $a=\frac{1}{1+z}$ ($a_0=1$). It is:
\begin{align*}
\frac{d\delta_m}{da} = -\frac{d\delta_m}{dz} (1+z)^2 \qquad
\frac{d^2\delta_m}{da^2} = \frac{d^2\delta_m}{dz^2}(1+z)^4 + 2\frac{d\delta_m}{dz} (1+z)^3
\end{align*}
\begin{align*}
\frac{d^2\delta_m}{dz^2}(1+z)^2  +
\frac{d\delta_m}{dz} (1+z) \left[\frac{1}{2}+ \frac{5C-6\Lambda (1+z)^2}{12H^2(1+z)^2}  \right] 
 -\left[\frac{3}{2} - \frac{5C+6\Lambda (1+z)^2}{12 H^2(1+z)^2}  \right ] \delta_m =0  
\end{align*}
At this point employ $C=-6H_0^2\Omega_{D}$ and $\Lambda =3H_0^2\Omega_\Lambda $:
\begin{align*}
0=\frac{d^2\delta_m}{dz^2}(1+z)^2  +
\frac{d\delta_m}{dz} (1+z)  \frac{(H/H_0)^2(1+z)^2-5 \Omega_{D} - 3 \Omega_\Lambda (1+z)^2}{2(H/H_0)^2(1+z)^2}  \\
 - \frac{3(H/H_0)^2(1+z)^2 + 5\Omega_{D}- 3\Omega_\Lambda (1+z)^2}{2 (H/H_0)^2(1+z)^2} \delta_m 
 \end{align*}
 and insert $H(z)/H_0$ given in \eqref{H2MLD}:
 \begin{align}
 0=\frac{d^2\delta_m}{dz^2}(1+z)^2  + \frac{1}{2}\frac{d\delta_m}{dz} (1+z)  \frac{\Omega_m(1+z)^5 - 2 \Omega_\Lambda (1+z)^2
 - 4 \Omega_{D} }{\Omega_m (1+z)^5 + \Omega_\Lambda (1+z)^2 +\Omega_{D} }  \label{DELTAALL}\\
 - \frac{\delta_m}{2} \frac{3\Omega_m(1+z)^5 +8 \Omega_{D}}{\Omega_m (1+z)^5 + \Omega_\Lambda (1+z)^2 +\Omega_{D}} 
 \nonumber 
\end{align}
Now we discuss various cases:\\

$\bullet$ When $\Omega_{D}=0$ and $\Omega_\Lambda =0$ we recover the GR evolution 
of the density contrast: 
\begin{align}
\delta_{m,GR} (z) = c_1(1+z)^{3/2}+ c_2 (1+z)^{-1} \label{DELTAGR}
\end{align}
The constants $c_1$ and $c_2$ are determined by initial conditions at a reference redshift: 
\begin{align*}
&\delta_{m,GR}(z_i) = c_1(1+z_i)^{3/2}+ c_2 (1+z_i)^{-1}\\
&\delta'_{m,GR}(z_i) = \tfrac{3}{2}c_1(1+z_i)^{1/2}- c_2 (1+z_i)^{-2} 
\end{align*}
The chosen reference is $z_i\gg1$ (matter dominated era). It is required that the fluctuation is small and the derivative is negative and small (initial growth of structures). This rules out $c_2$ as unphysical and poses the``adiabatic condition''
\begin{align}
\delta'_m (z_i) = - \frac{\delta_m(z_i)}{1+z_i} \label{eq:adiabatic condition}
\end{align}
that is used as a criterion to fix the constants in more general conditions.\\
\quad\\

$\bullet $ Eq.\eqref{DELTAALL} is now studied by assuming $\Omega_\Lambda \neq 0$, and $\Omega_D$ negligible. 
This is standard $\Lambda$CDM (see Peebles \cite{Peebles}, Martel \cite{Martel91}). In this approximation
we put $\alpha = \Omega_{m}/\Omega_\Lambda $. The common acceptance is $\alpha\approx 3/7$ if $\Omega_{D}=0$.
The equation is:
 \begin{align}
 0=\frac{d^2\delta_m}{dz^2}(1+z)^2  + \frac{1}{2}\frac{d\delta_m}{dz} (1+z)  \frac{\alpha (1+z)^3 - 2}{\alpha (1+z)^3 +1} 
 - \frac{\delta_m}{2} \frac{3\alpha (1+z)^3 }{\alpha (1+z)^3 +1} 
 \nonumber 
\end{align}
Let us introduce the new variable $ x=\frac{1}{\alpha} (1+z)^{-3}$. 
The differential equation becomes
\begin{align*}
 0=x^2(1+x)\frac{d^2\delta_m}{dx^2}  + \frac{x}{6}\frac{d\delta_m}{dx} (10x+7)  -\frac{1}{6}\delta_m
\end{align*}
It reduces to a hypergeometric equation by appropriate $\gamma $ in $\delta_m(x) = x^\gamma F(x)$. \\
$F$ solves the equation
$$ x^2(1+x)F'' +\frac{x}{6} F' [2x(5+6\gamma)+7+12\gamma] +\frac{F}{6}(6\gamma^2+6\gamma -1  +2x\gamma (3\gamma +2)]=0$$
The choices $\gamma = \frac{1}{3}$ and $\gamma=-\frac{1}{2}$ give two hypergeometric equations for $G(x)=F(-x)$:
$$ x(1-x)G'' + \frac{G'}{6} [-2x(5+6\gamma )+7+12\gamma ] - \frac{G}{3}\gamma (3\gamma +2) =0$$
For $\gamma =-\frac{1}{2}$: $G(x)= {}_2F_1(\frac{1}{6}, -\frac{1}{2}; \frac{1}{6}; x) = \sqrt{1-x}$. 
For $\gamma =\frac{1}{3}$: $G(x) = {}_2F_1 (1,\frac{1}{3}; \frac{11}{6}; x)$. \\
The general solution is:
\begin{align*}
\delta_m (x) = \kappa_1 \sqrt{\frac{1+x}{x}} + \kappa_2 \, x^{1/3} {}_2F_1 \left (1,\frac{1}{3}; \frac{11}{6}; -x \right )
 \end{align*}
It coincides with the case $m=0$ for the evolution of density perturbations in Newtonian approximation with
$\Lambda = 3\alpha a(t)^{-m}$, by Silveira and Waga \cite{Silveira94}.\\
Back to redshift:
\begin{align}
\delta_m (z) = \kappa_1 \sqrt{1+\alpha (1+z)^3} +\kappa_2 \frac{1}{1+z}\;
{}_2F_1\left (1,\frac{1}{3}; \frac{11}{6}; -\frac{1}{\alpha (1+z)^3} \right )
\nonumber
\end{align}
In the dominant matter regime, $\alpha (1+z)^3$ is large and the hypergeometric function is around unity. The 
$\Lambda$CDM density contrast is given by the above formula with $\kappa_1=0$ and becomes GR for large $z$.
For large $z_i$ it fulfills the adiabatic condition \eqref{eq:adiabatic condition}.
\quad

\begin{figure}[t]
\begin{center}
\includegraphics*[width=10cm,clip=]{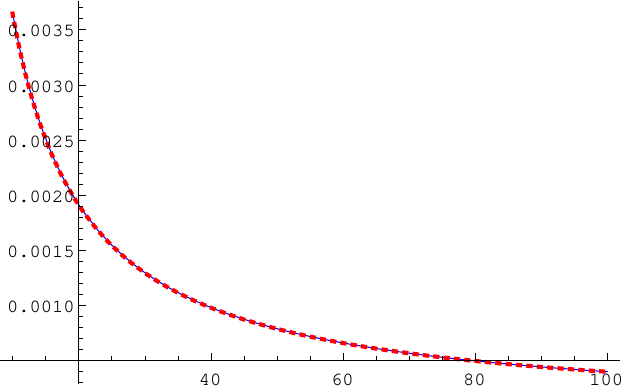}
\caption{\label{delta} The density contrast $\delta_m(z)$ with initial conditions $\delta_m(400)=0.0001$
and $\delta'_m(400)=-\delta_m(400)/401$ for: 
$\Lambda$CDM (red, dotted), 
CKG (blue) with CC+BAO data.
The blue line is numerical (Mathematica).}
\end{center}
\end{figure}

$\bullet $ Eq.\eqref{DELTAALL} is studied with $\Omega_\Lambda =0$ (in CKG the cosmological constant $\Lambda$ is an integration constant, not a term of the field equations). Now $\Omega_{D}$ comes into play.\\
 Let $\beta =\Omega_{m}/\Omega_{D}$. 
\begin{equation}
\frac{d^2\delta_m}{dz^2}(1+z)^2  +\frac{1}{2} \frac{d\delta_m}{dz} (1+z) \frac{\beta\left(1+z\right)^5-4}{\beta\left(1+z\right)^5+1}  
 - \frac{1}{2} \frac{3\beta\left(1+z\right)^5+8}{\beta\left(1+z\right)^5+1}   \delta_m =0  \label{eq:redshift final}
\end{equation}
The equation is solved in the new variable 
$x=\dfrac{1}{\beta(1+z)^{5}} $. Then 
$(1+z)\dfrac{d\delta_m}{dz}=-5x\dfrac{d\delta_m}{dx}$ and  
$(1+z)^2 \dfrac{d^2\delta_m}{dz^2}=30x\dfrac{d\delta_m}{dx} +25x^2 \dfrac{d^2\delta_m}{dx^2}$.
Equation (\ref{eq:redshift final}) takes the form 
\begin{equation}
x^2(1+x)\frac{d^2 \delta_m}{dx^2} + \frac{x}{10} (16 x+11)\frac{d\delta_m}{dx} - \frac{8x+3}{50}\delta_m = 0 
\label{eq:density contrast versus x}
\end{equation}
Now  change the dependent variable $\delta_m (x)= x^\gamma F(x)$:
%
\begin{align*}
x(1+x)\frac{d^2 F}{dx^2}&  + \frac{dF}{dx} \left [2\gamma(1+x) + \frac{16x+11}{10} \right ] \\
&+\left(\gamma-\frac{1}{5}\right) \left [\frac{1}{x}\left(\gamma +\frac{3}{10}\right)  +\left(\gamma+\frac{4}{5}\right)
 \right ] F =0
\end{align*}
The value $\gamma = \frac{1}{5} $ simplifies the equation:  $x(1+x)F''  + F' \left (2x+\frac{3}{2}\right ) =0$ i.e.
$$ \frac{d}{dx} [ x(1+x) F'(x) +\tfrac{1}{2} F(x)]= 0$$
Then: $2x(1+x)F'(x) + F(x) = f_1$ where $f_1$ is a constant, 
with solution $ F(x) = f_2  \sqrt{\dfrac{1+x}{x}} + f_1 $. The final result is a simple expression:
\begin{align}
\delta_m(z) = f_1 \frac{1}{1+z} + f_2 \frac{\sqrt{1+\beta (1+z)^5}}{1+z} \label{DELTASOLUTION}
\end{align}
%
For large $\beta (1+z)$ (matter dominated regime) we recover the functional form of the GR result \eqref{DELTAGR}.

The evolution of the matter and of the dark energy densities  are
$$\Omega_m(z)=
\dfrac{\Omega_m (1+z)^{3}}{(H(z)/H_0)^2}, \qquad \Omega_D(z)=\dfrac{\Omega_D(1+z)^{-2}}{(H(z)/H_0)^2}$$
Thus the condition of matter dominance $\Omega_{m}(z)/\Omega_{D}(z)>1$ is 
$\beta(1+z)^5>1$, i.e. $z>\sqrt[5]{\Omega_D/\Omega_m}-1$.
Since for large $z$ the expression of the density contrast 
is very similar to the GR expression, we apply the
adiabatic condition (\ref{eq:adiabatic condition}), as usually done in the literature:
\begin{equation*}
\dfrac{d\delta_{m}(z)}{dz}= -\dfrac{f_1}{(1+z)^2} + f_2 \dfrac{3\beta(1+z)^{5}-2}{2(1+z)^{2}\sqrt{1+\beta(1+z)^{5}}}
\end{equation*}
The adiabatic condition (\ref{eq:adiabatic condition}) at large $z_i$ imposes $f_2=0$.
The density contrast 
grows linearly in the scale factor 
as in GR.\\

$\bullet$ Evolution in CKG with $\Omega_\Lambda \neq 0$, $\Omega_m+\Omega_\Lambda +\Omega_D=1$.\\
Eq.\eqref{DELTAALL} is rewritten in the variable $x$:
$$x=\frac{1}{\beta (1+z)^5},\qquad 
\beta = \frac{\Omega_m}{\Omega_D},\qquad 
 \eta = \sqrt[5]{ \frac{\Omega^5_\Lambda}{\Omega_m^2\Omega^3_D} } $$
\begin{align}
 0=25 x^2 \frac{d^2\delta_m}{dx^2}  + \frac{5}{2}x\frac{d\delta_m}{dx} \frac{11 +14 \eta x^{3/5}+16x}{1+\eta x^{3/5}+x}
 - \frac{\delta_m}{2} \frac{3+8x}{1+\eta x^{3/5}+x} \label{44}
\end{align}
For $z$ in the range 10-100 (small $x$) the leading behaviour of the independent solutions is read in the approximation
$ 0=25 x^2 \frac{d^2\delta_m}{dx^2}  + \frac{55}{2}x\frac{d\delta_m}{dx}  - \frac{3\delta_m}{2} $ that gives an expanding
and a decreasing asymptotic solution: $\delta_m (x) = \kappa_- x^{-3/10} +\kappa_+ x^{1/5}$.

Starting with the factorization $\delta_{m}^{-}=x^{-3/10}F^{-}(x) $ 
eq.\eqref{44} becomes
\begin{align*}
25x^2 \frac{d^2 F^{-}}{dx^2}(1+\eta x^{3/5}+x)+\frac{5}{2} x [5+10x+8\eta x^{3/5}] \frac{dF^{-}}{dx}-\frac{1}{4}F[3\eta x^{3/5}+25x]=0
\end{align*}
Now make the ansatz $F^{-}(x)=\sqrt{1+ A\eta x^{3/5}+Bx}$. Neglecting terms of order $x^{6/5}$ and smaller,
it is: $x^{2}\dfrac{d^{2}F^{-}}{dx^{2}}=-\dfrac{3}{25}A\eta x^{3/5}$ and $x\dfrac{dF^{-}}{dx}=\dfrac{3}{10} A\eta x^{3/5}+\dfrac{B}{2}x$. The equation for $F^{-}$ simplifies to $3(A-1)\eta x^{3/5}+25x(B-1)=0$
that gives $A=B=1$.
One then obtains the following expression, correct up to vanishing terms $x^{6/5}$.
\begin{align}
\delta_m^{-} (x) = \sqrt{\frac{1+\eta x^{3/5} +x}{x^{3/5}} } \label{DELTAM} 
\end{align}
Next, by posing $\delta^+(x)= \delta^-(x)F^+(x)$ one obtains the other approximate
solution (this procedure is equivalent to exploiting the Wronskian).
\begin{align}
&\delta_m^{+} (x) =  \sqrt{\frac{1+\eta x^{3/5} +x}{x^{3/5}}} \int^x_0 \frac{dx'}{\sqrt{x'} (1+\eta x'^{3/5}+x')^{3/2}} \label{DELTAP}
\end{align}
Similar expressions, with different exponents, were obtained by Hugo Martel in the study
of density perturbations of the Friedmann equations with the $\Lambda $ term in linear adiabatic regime (eqs. 10 and 11 in \cite{Martel91}). 

Eq.\eqref{DELTAALL} is also solved numerically with {\sf NDSolve} (Mathematica7). 
For comparison with other approximations, the initial conditions $\delta_m(z_i)$ and $\delta'_m(z_i)$ are chosen 
identical. 
We put $\delta_m(400)=0.0001$ 
and $\delta_m'(400) = -\delta_m(400)/401$.\\
Despite the different analytic expressions and values of $\Omega_j $, the plot of CKG gravity with or without cosmological constant does not show significative difference with the expanding mode of GR and  $\Lambda$CDM in
the matter-dominated phase.

\section{Conclusions}
A spacetime is generalized RW  if and only if it admits a divergence-free, acceleration free, perfect fluid Sinyukov-like tensor. Such tensors are conformal Killing tensors, and are candidates for the dark sector of Conformal Killing Gravity.\\
 We study the Friedmann equations for CKG  and determine the analytic form of $H(z)$. Neglecting 
radiation, in spatially flat RW spacetime, $H(z)$ is fitted against experimental datasets
for CC or CC+BAO to estimate the dark sector parameters $\Omega_D$, $\Omega_\Lambda$ and the matter parameter
$\Omega_m$. While the model shows non relevant deviation of $H(z)$ in CKG from $\Lambda$CDM  in the past, the future is driven by the dark term to a future singularity: a big rip if $\Omega_D$ and $\Omega_\Lambda$ are
positive, an exotic one if $\Omega_D<0$.\\
Next we solve the equation for the evolution of the density contrast $\delta_m(z)$ in the linear regime.
The  solution in CKG, for values $\Omega_m+\Omega_\Lambda+\Omega_D=1$
shows no sensible deviation in the matter dominated-regime from the $\Lambda$CDM or GR solutions.

While Conformal Killing gravity, as an extension of Einstein's gravity, does not contradict the present view of the past evolution of
the universe, it differs in describing the late accelerated phase. New data for $H(z)$ are needed to solve the sign ambiguity of $\Omega_D$ that influences the evolution picture offered by the theory.

\section{Appendix 1: proof of theorem \ref{PFCKT}}
The perfect fluid tensor $Ag_{jk}+Bu_{j}u_{k}$ is a CKT if (\ref{CKT}) holds, with conformal vector (\ref{ETAVECTOR}):
\begin{align}
(n+2)\eta_{i}=(n+2)\nabla_{i}A-\nabla_{i}B+2\dot{B}u_{i}+2B\dot{u}_{i}+2Bu_{i}\nabla_{j}u^{j}\label{ETAALPHA}
\end{align}
The condition (\ref{CKT}) with (\ref{ETAALPHA}) is 
\begin{align}
0= & (n+2)[\nabla_{i}(Bu_{j}u_{l})+\nabla_{j}(Bu_{i}u_{l})+\nabla_{l}(Bu_{i}u_{j})]\label{CKTPF}\\
 & -g_{jl}[-\nabla_{i}B+2\dot{B}u_{i}+2B\dot{u}_{i}+2Bu_{i}(\nabla_{p}u^{p})]\nonumber \\
 & -g_{il}[-\nabla_{j}B+2\dot{B}u_{j}+2B\dot{u}_{j}+2Bu_{j}(\nabla_{p}u^{p})]\nonumber \\
 & -g_{ij}[-\nabla_{l}B+2\dot{B}u_{l}+2B\dot{u}_{l}+2Bu_{l}(\nabla_{p}u^{p})]\nonumber 
\end{align}
Contraction with $u^{i}u^{j}u^{l}$ is $0=3(n+2)u^{j}u^{l}u^{i}\nabla_{i}(Bu_{j}u_{l})-3[3\dot{B}+2B(\nabla_{p}u^{p})]$, i.e.
\begin{align}
\frac{\nabla^{p}u_{p}}{n-1}=H=\frac{\dot{B}}{2B}\label{EXPAN}
\end{align}
Contraction of (\ref{CKTPF}) with $u^{j}u^{l}$: 
\begin{align*}
0= & (n+2)[\nabla_{i}B-2\dot{B}u_{i}-2B\dot{u}_{i}]\\
 & -\nabla_{i}B+2\dot{B}u_{i}+2B\dot{u}_{i}+2Bu_{i}(\nabla_{p}u^{p})-2u_{i}[-3\dot{B}-2B(\nabla_{p}u^{p})]\\
= & (n+1)\nabla_{i}B-2(n-2)\dot{B}u_{i}-2(n+1)B\dot{u}_{i}+6Bu_{i}(\nabla_{p}u^{p})
\end{align*}
Use (\ref{EXPAN}) and obtain eq.(\ref{nablab}): $\nabla_{i}B=-\dot B u_{i}+2B\dot{u}_{i}$.
This and (\ref{EXPAN}) are inserted in (\ref{CKTPF}): 
\begin{align*}
0=2B(\dot{u}_{i}u_{j}u_{l}+u_{i}\dot{u}_{j}u_{l}+u_{i}u_{j}\dot{u}_{l})+B[\nabla_{i}(u_{j}u_{l})+\nabla_{j}(u_{i}u_{l})+\nabla_{l}(u_{i}u_{j})]\\
-(u_{j}u_{l}+g_{jl})\dot{B}u_{i}-(u_{i}u_{l}+g_{il})\dot{B}u_{j}-(u_{i}u_{j}+g_{ij})\dot{B}u_{l}
\end{align*}
Contraction with $u^{i}$: $\nabla_j u_l +\nabla_l u_j =\frac{\dot{B}}{B}(u_{j}u_{l}+g_{jl})-u_{j}\dot{u}_{l}-u_{l}\dot{u}_{j}$.
If we insert the standard decomposition \eqref{CANONICAL} of $\nabla_j u_l$, the equation is satisfied with shear 
$\sigma_{il}=0$. The CKT condition
(\ref{CKT}) is now identically verified. \\
The conformal vector (\ref{ETAVECTOR}) becomes $(n+2)(\eta_{i}-\nabla_{i}A)=3\dot{B}u_{i}+2B\nabla_{i}u^{i}$
i.e. $\eta_{i}=\nabla_{i}A+\dot{B}u_{i}$.

Let us prove the opposite. Suppose that the perfect
fluid tensor (\ref{eq:perfect fluid tensor}) has shear-free velocity
with expansion $\dot{B}/2B$ and $B$ that satisfies (\ref{nablab}).\\
$\nabla_{i}K_{jl}=g_{jl}\nabla_{i}A+(-\dot{B}u_{i}+2B\dot{u}_{i})u_{j}u_{l}+B(u_{j}\nabla_{i}u_{l}+u_{l}\nabla_{i}u_{j})=g_{jl}\nabla_{i}A+B(2\dot{u}_{i}u_{j}u_{l}-u_{i}u_{j}\dot{u}_{l}-u_{i}\dot{u}_{j}u_{l})+\frac{1}{2}\dot{B}(u_{j}g_{il}+u_{l}g_{ij})+\frac{1}{2}B(u_{j}\omega_{il}+u_{l}\omega_{ij}).$\\
In the cyclic sum many terms cancel: $\nabla_{i}K_{jl}+$
cyclic $=g_{jl}(\nabla_{i}A+\dot{B}u_{i})+$ cyclic. The CKT condition
is satisfied with $\eta_i = \nabla_i A + \dot B  u_i$. \hfill $\square$

\vfill
\section{Appendix 2: CC and BAO datasets for $H(z)$ \cite{Sharov18,Tomasetti23}.}
\begin{center}
\begin{tabular}{| l | c | c || c | c| c|}
\hline
{\quad}z & H(z) & $\sigma$ & z & H(z) & $\sigma $\\
\hline\hline
0.070 & 69 & 19.6 & 0.24 & 79.69 & 2.99 \\
0.090 & 69 & 12 & 0.30 & 81.7 & 6.22 \\ 
0.120 & 68.6 & 26.2 & 0.31 & 78.18 & 4.74\\
0.170 & 83 & 8  & 0.34 & 83.8 & 3.66\\
0.1791 & 75 & 4 & 0.35 & 82.7 & 9.1\\
0.1993 & 75 & 5 & 0.36 & 79.94 & 3.38\\ 
0.200 & 72.9 & 29.6 & 0.38 & 81.5 & 1.9\\
0.270 & 77 & 14 & 0.40 & 82.04 & 2.03\\
0.280 & 88.8 & 36.6 & 0.43 & 86.45 & 3.97\\ 
0.3519 & 83 & 14 & 0.44 & 82.6 & 7.8\\
0.3802 & 83 & 13.5 & 0.44 &  84.81 & 1.83\\  
0.400 & 95 & 17 & 0.48 & 87.79 & 2.03\\
0.4004 & 77 & 10.2 & 0.51 & 90.4 & 1.9\\ 
0.4247 & 87.1 & 11.2 & 0.52 & 94.35 & 2.64\\
0.4497 & 92.8 &12.9 &0.56 & 93.34 & 2.3\\
0.470 & 89 & 34 & 0.57 & 87.6 &  7.8\\
0.4783 & 80.9& 9 & 0.57 & 96.8 & 3.4\\
0.480& 97& 62 &0.59 & 98.48 & 3.18\\ 
0.593& 104& 13 & 0.60 & 87.9 & 6.1\\
 0.6797& 92& 8 &0.61 & 97.3 & 2.1\\
 0.7812& 105& 12 & 0.64 & 98.82 & 2.98\\ 
 0.8754& 125& 17 & 0.73 &97.3 & 7.0\\
 0.880& 90& 40 & 2.30 & 224 & 8.6\\
 0.900& 117& 23 & 2.33 & 224 & 8\\ 
 1.037& 154& 20 &  2.34 & 222 & 8.5\\ 
 1.26 & 135 & 65 & 2.36 & 226 & 9.3\\
 1.300& 168& 17 &  && \\
 1.363& 160& 33.6 &  & & \\
1.430&177& 18 & & & \\
1.530& 140& 14 & & &\\
1.750& 202& 40 & & &\\
1.965& 186.5& 50.4 & & & \\
\hline
\end{tabular}
\end{center}


\begin{thebibliography}{1}
%
\bibitem{Abramo07}
L.~R.~Abramo, R.~C.~Batista, L.~Liberato and R.~Rosenfeld,
{\em Structure formation in the presence of dark energy perturbations},
JCAP {\bf 11} (2007) 012 (21pp).
%
%
\bibitem{Barnes23a}
A.~Barnes, 
{\em Vacuum static spherically symmetric spacetimes in Harada's theory}, 
(2023), arXiv:2309.05336 [gr-qc],  

\bibitem{Barnes23b}
A.~Barnes, {\em Harada-Maxwell static spherically symmetric spacetimes}, (2023), 
arXiv:2311.09171 [gr.qc], 
%
\bibitem{Barnes24}
A.~Barnes, {\em pp-waves in conformal Killing gravity}, (2024), 
arXiv:2404.09310 [gr.qc], 
%
\bibitem{Cai16}
Y.~F.~Cai, S.~Capozziello, M.~De Laurentis and E.~N.~Saridakis,
\textit{$f(T)$ teleparallel gravity and cosmology}, 
Rep. Prog. Phys. \textbf{79} n.10, (2016) 106901.
%
\bibitem{Caldwell02}
R.~R.~Caldwell,
\textit{A phantom menace? Cosmological consequences of a dark energy component with super-negative equation of state}, Phys. Lett. B {\bf 545} (2002) 23--29.
%
\bibitem{Caldwell03}
R.~R.~Caldwell, M.~Kamionkowski, and N.~N.~Weinberg,  
\textit{Phantom Energy: Dark Energy with $w<-1$ causes a Cosmic Doomsday}, 
Phys. Rev. Lett. {\bf 91}n (2003) 071301.
%
\bibitem{Capozziello11}
S.~Capozziello and M.~De~Laurentis, 
\textit{Extended theories of gravity}, 
Phys. Rep. {\bf 509} n.4--5 (2011) 167.
%
\bibitem{Capozziello 22}
S.~Capozziello, C.~A.~Mantica, and L.~G.~Molinari,
\textit{Geometric perfect fluids from Extended Gravity}, 
Europhys. Lett. \textbf{137} (2022), 19001 (7pp). 
%
\bibitem{Chamseddine13}
A.~H.~Chamseddine, and V.~Mukhanov,
\textit{Mimetic dark matter}, 
JHEP \textbf{11} (2013), 135.
%
\bibitem{Clement24}
G.~Clem\'ent and K.~Nouicer,
{\em Spherical symmetric solutions of conformal Killing gravity: black holes, wormholes and sourceless cosmologies}, 
(2024) arXiv:2404.00328v1 [gr-qc],
%
\bibitem{Coll06}
B.~Coll, J.~J.~Ferrando, J.~A.~S\'aez, 
\textit{On the geometry of Killing and conformal tensors}, 
J. Math. Phys. {\bf 47}, 062503 (2006). 
%
\bibitem{Copeland06}
E.~J.~Copeland, M.~Sami, and S.~Tdujikawa, 
\textit{Dynamics of dark energy}, 
Int. J. Mod. Phys. D {\bf 15} (2003) 1753.
%
\bibitem{deHaro23}
J.~de Haro, S.~Nojiri, S.~D.~Odintsov, V.~K.~Oikonomou, and A.~Pan, 
\textit{Finite-time cosmological singularities and the possible fate of the Universe}, 
Phys. Rep. {\bf 1034} (2023) 1--114.
%
\bibitem{Farsi22}
B.~Farsi and A.~Sheykhi,
{\em Structure formation in mimetic gravity},
Phys. Rev. D {\bf 106} 024053 (2022). 
%
\bibitem{Farsi23}
B.~Farsi, A.~Sheykhi, and M.~Khodadi, 
{\em Evolution of spherical overdensities in energy-momentum-squared gravity},
Phys. Rev. D {\bf 108} (2023) 023524. 
%
\bibitem{Formella 95} 
S.~Formella, 
\textit{On some class of of nearly conformally symmetric manifolds}, 
Colloq. Math. \textbf{68} n.1 (1995) 149--164. 
%
\bibitem{Gangopadhyay23}
M.~R.~Gangopadhyay, M.~Sami, and M.~K.~Sharma, 
\textit{Phantom dark energy as a natural selection of evolutionary processes aˆ la genetic algorithm and cosmological tensions}, Phys. Rev. D {\bf 108} (2023) 103526 (13pp).
%
\bibitem{Junior24}
J.~T.~S.~S.~Junior, F.~S.~N.~Lobo and M.~E.~Rodriguez,
{\em (Regular) black holes in conformal Killing gravity coupled to nonlinear electrodynamics and scalar fields}
Class. Quantum Grav. {\bf 41} (5) (2024) 055012 (24pp).  
%
\bibitem{Junior24b}
 J.~T.~S.~S.~Junior, F.~S.~N.~Lobo, M.~E.~Rodrigues,  
\textit{Black bounces in conformal Killing gravity}, 
(2024), arXiv:2404.09702
%
\bibitem{Harada23a} 
J.~Harada,
\textit{Gravity at cosmological distances: explaining the accelerating expansion without dark energy},
Phys. Rev. D \textbf{108} (4) (2023) 044031. 
%
\bibitem{Harada23b} 
J.~Harada,
\textit{Dark energy in conformal Killing gravity}, 
Phys. Rev. D \textbf{108} (2023) 104037.
%
\bibitem{Kamenshick01}
A.~Kamenshchik, U.~Moschella, V.~Pasquier, {\em An alternative to quintessence}, 
Phys. Lett. B {\bf 511} (2001) 265--268.
%
\bibitem{Mantica12}
C.~A.~Mantica and L.~G.~Molinari, 
\textit{Riemann compatible tensors}, 
Colloq. Math. \textbf{128} n.2 (2012) 197--210. 
%
\bibitem{Mantica16}
C.~A.~Mantica and L.~G.~Molinari, 
\textit{On the Weyl and Ricci tensors in Generalized Robertson-Walker spacetimes}, 
J. Math. Phys. \textbf{57} (2016) 102502 (6pp). 
%
\bibitem{Mantica 17}
C.~A.~Mantica and L.~G.~Molinari, 
\textit{Generalized Robertson-Walker spacetimes, a survey}, 
Int. J. Geom. Meth. Mod. Phys. \textbf{14} n.3 (2017) 1730001.
%
\bibitem{Mantica 23 a} 
C.~A.~Mantica and L.~G.~Molinari, 
\textit{Note on Harada's conformal Killing gravity}, 
Phys. Rev. D {\bf 108} (2023) 124029 (5pp). 
%
\bibitem{Mantica 24} 
C.~A.~Mantica and L.~G.~Molinari, 
\textit{Friedmann equations in the Codazzi parametrization of Cotton and extended theories of gravity and the dark sector},
Phys. Rev. D {\bf 109} (2024) 044059 (12 pp). 
%
\bibitem{Martel91}
H.~Martel, 
{\em Linear perturbation theory and spherical overdensities in $\Lambda\neq 0$ Friedmann model},
Astrophys. J. {\bf 377} (1991) 7--13. 
%
\bibitem{Moresco12} 
M.~Moresco et al. , {\em Improved constraints on the expansion rate of the Universe up to $z\approx 1.1$ from the spectroscopic evolution of cosmic chronometers}, JCAP {\bf 08} 2012.

\bibitem{Moresco23}
M.~Moresco, 
{\em Addressing the Hubble tension withcosmic chronometers}, 
arXiv:2307.09501 [stro-ph.CO];
%
\bibitem{Nojiri05a}
S.~Nojiri and S.~D.~Odintsov,
\textit{Modified Gauss Bonnet theory as gravitational alternative for dark energy}, 
Phys. Lett. B \textbf{631} (1-2) (2005) 1--6.
%
\bibitem{Nojiri05}
S.~Nojiri, S.~D.~Odintsov, S.~Tsujikawa, \textit{Properties of singularities in (phantom) dark energy universe},
Phys. Rev. D {\bf 71} (2005) 063004. 
%
\bibitem{Nojiri17}
S.~Nojiri, S.~D.~Odintsov, and V.~K.~Oikonomou,
\textit{Modified Gravity Theories on a Nutshell: inflation, bounce and late-time evolution}, 
Phys. Rep. B \textbf{692} (2016), 1--104.
%
\bibitem{Odintsov23}
S.~D.~Odintsov, D.~S-C.~G\'omez and G.~S.~Sharov, 
\textit{Exponential F(R) gravity with axion dark matter}, 
Phys. Dark Universe {\bf 42} (2023) 101369.
%
\bibitem{Qi23}
Jing-Zhao Qi, Ping Meng, Jing-Fei Zhang, and Xin Zhang, 
{\em Model-independent measurement of cosmic curvature with the latest $H(z)$ and SNe Ia data: A comprehensive investigation},  
Phys. Rev. D {\bf 108} (2023) 063522 (10pp).
%
\bibitem{Peebles}
P.~Peebles, The large-scale structure of the Universe, Princeton University Press, 1980.
%
\bibitem{Rani 03}
R.~Rani, S.~B.~Edgar and A.~Barnes, 
\textit{Killing tensors and conformal Killing tensors from conformal Killing vectors}, 
Class. Quantum Grav. \textbf{20} (2003), 1929--1942.
%
\bibitem{Ratra88}
B.~Ratra and P.~J.~E.~Peebles,
\textit{Cosmological consequences of a rolling homogeneous scalar field}, 
Phys. Rev. D {\bf 37} n.12 (1988) 3406--3427.
%
\bibitem{Saridakis21}
E.~N.~Saridakis, R.~Lazkoz, V.~Salzano, P.~V.~Monitz, S.~Capozziello, J.~B.~Jim\'enez, M.~De Laurentis, and G.~J.~Olmo, Modified gravity and cosmology. An update by the CANTATA network. Springer, 2021.
%
 \bibitem{Sharma10}
R.~Sharma and A.~Ghosh, 
\textit{Perfect fluid space-times whose energy-momentum tensor is conformal Killing}, 
J. Math. Phys. {\bf 51}, 022504 (2010). 
%
\bibitem{Sharov18}
G.~S.~Sharov and V.~O.~Vasiliev, 
{\em How predictions of cosmological models depend on Hubble parameter datasets},
Math. Modelling and Geom. {\bf 6} n.1 (2018) 1--20.
%
\bibitem{Silveira94}
D.~Silveira and I.~Waga, 
{\em Decaying $\Lambda $ cosmologies and power spectrum}, 
Phys. Rev. D {\bf 50} n.8 (1994) 4890--4894. 
%
\bibitem{Sotiriu10} 
T.~P.~Sotiriu and V.~Faraoni, 
\textit{$f(R)$ theories of gravity}, 
Rev. Mod. Phys. \textbf{82}, 451 (2010).
%
\bibitem{Step12}
S.~Stepanov and J.~Mike$\check{s}$, 
\textit{Seven invariant classes of the Einstein equations and projective mappings}, 
AIP Conf. Proc. \textbf{1460} (2012), 221--225. 
%
\bibitem{Tomasetti23}
E.~Tomasetti, M.~Moresco, N.~Borghi, K.~Jiao, A.~Cimatti, L.~Pozzetti, A.~C.~Carnall, R.~J.~McLure, and L.~Pentericci,
{\em A new measurement of the expansion history of the Universe at z = 1.26 with cosmic chronometers in VANDELS},
Astronomy \& Astrophysics {\bf 679} (2023) A96 (18pp).
%
\bibitem{Usman23}
M.~Usman and A.~Jawad,
{\em Matter growth perturbations and cosmography in modified torsion cosmology},
Eur. Phys. J. C {\bf 83}:958 (2023), (11pp).
%
\bibitem{Ziaie20}
A.~H.~Ziaie, H.~Moradpour and H.~Shabani,
{\em Structure formation in Generalized Rastall gravity}
Europ. Phys. J. Plus {\bf 135} 916 (2020). 
%
\end{thebibliography}
\end{document}